\documentclass[12pt,preprint]{aastex} 
\slugcomment{Submitted to Astrophysical Journal} 

\shorttitle{Luminous O-Rich SNR in NGC 4449} 
\shortauthors{Patnaude \& Fesen} 

\begin{document} 

\title{{\it Chandra} Observations of the Luminous, O-Rich
SNR in the Irregular Galaxy NGC 4449}

\author{Daniel J.\ Patnaude \& Robert A.\ Fesen }
\affil{6127 Wilder Laboratory, Physics \& Astronomy Department \\
       Dartmouth College, Hanover, NH 03755}

\begin{abstract} 

An analysis of a 29 ksec {\it Chandra} ACIS-S
observations of the young, Cassiopeia A-like supernova remnant in the 
irregular galaxy NGC 4449 is presented. The observed $0.5 - 2.1$ keV 
spectrum reveals the likely presence of several emission lines including \ion{O}{8} 
at 0.65 and 0.77 keV, \ion{Ne}{10} at 1.05 keV, 
\ion{Mg}{11} at 1.5 keV, and \ion{Si}{13} at 1.85 keV. 
From the observed spectrum, we derive 
an N$_{\rm H}$ = 1 $\times$ $10^{21}$ cm$^{-2}$ and 
an X-ray temperature of T $\approx$ 9 $\times$ 10$^{6}$ K.
A non-equilibrium ionization
fit to the spectrum suggests an overabundance of oxygen around 20 times
solar, consistent with the remnant's UV and optical emission-line properties.
We discuss the remnant's approximate X-ray derived elemental abundances and compare its 
X-ray spectrum and luminosity to other oxygen-rich remnants.

\end{abstract} 
\keywords{galaxies: individual -- nebulae: supernova remnants -- X-rays: 
sources} 

\section{INTRODUCTION} 

At present, there are only about a half-dozen O-rich supernova remnants 
(SNRs) known. They represent the  remains of high-mass
stars (M $\gtrsim$ 15 M$_{\sun}$) and thus are especially useful for studying  
the elemental abundances of explosive nucleosynthesis
processes in core-collapse supernovae. The younger the remnant is,
the more likely its properties accurately reflect true SN ejecta abundances  
unaffected by dilution from swept-up circumstellar and interstellar material.

Currently, the youngest Galactic remnant is the
$\simeq$ 320 yr old O-rich SNR Cassiopeia A (Cas~A). 
However, with an estimated age of around 100 yr (\citealt{blair83,blair98}), 
the O-rich SNR located in the Magellanic type irregular galaxy NGC 4449
represents the youngest example of this type of remnant and is therefore
an object of particular interest.

The NGC 4449 SNR was first identified  
by \citet{seaquist78}, who found a strong and point-like,
non-thermal radio source approximately 1\arcmin~north
of the nucleus of this galaxy 
coincident with an \ion{H}{2} 
region cataloged by \citet{sabbadin79}. 
Optical spectrophotometry of
this source by \citet{balick78} showed two components: (1) narrow lines 
belonging to a conventional \ion{H}{2} region, and (2) broad 
[\ion{O}{3}] $\lambda \lambda$4959,5007 and 
[\ion{O}{1}] $\lambda \lambda$6300,6363 line emissions 
which they attributed to a young
SNR similar to Cas A.

Optical and UV spectra taken by \citet{kirshner80}, \citet{blair83}, and \citet{blair84} 
confirmed and extended this
interpretation and yielded a rough analysis of the physical conditions
in the SNR and the associated \ion{H}{2} region. They found
optical line widths implying an expansion velocity of 
3500 km s$^{-1}$ and 
emphasized the extremely low abundance of hydrogen in the ejecta. 
Furthermore, they estimated that the total mass of oxygen
in the optical was 10$^{-2}$ $M_{\sun}$, 
some 50 times higher than the value 
derived for Cas A \citep{peimbert71}, and estimate a progenitor mass 
$\approx$ 30 $M_{\sun}$.

Using data from the {\it Einstein X-Ray Observatory} High Resolution Imager, 
\citet{blair83} found 
the SNR to also be highly luminous in X-rays,
with L$_{\rm x}$ $\approx$ 10$^{39}$ erg s$^{-1}$
assuming a distance of 5.0 Mpc.  Even at a smaller estimated distance
of 3.9 Mpc \citep{devaucouleurs91,hunter99}, this SNR is nearly
10 times more luminous in X-rays than Cas A, 
the most luminous young SNR in our Galaxy, and several times  
more luminous than the bright O-rich SNR 
N132D in the Large Magellanic Cloud (L$_{\rm x}$ $\approx$ 6 $\times$ 
10$^{37}$ erg s$^{-1}$; \citealt{hughes87}).

More recently, \citet{vogler97} studied 
{\it ROSAT} ~PSPC and HRI detected NGC~4449
sources, including the SNR. They were able to fit the remnant's spectrum
to an absorbed thermal bremsstrahlung model, reporting a lower 
limit on both the absorbing
column $N_H \geq$ 7 $\times$ 10$^{20}$ cm$^{-2}$ and temperature of
$\geq$ 600 eV corresponding to T $\geq$ 7 $\times$ 10$^{6}$ K.  
They also found a L$_{\rm x}$ of $\approx$ 1.8 $\times$ 10$^{38}$
erg s$^{-1}$, a value about 50\% that of the Blair et al. (1983)
{\it Einstein} 1978 estimate.

Here we report on observations of the SNR in NGC 4449 obtained with the
{\it Chandra X-ray Observatory} \citep{weisskopf96}. 
In \S \ref{sec:obs} we describe the observations
and data reduction. In \S \ref{sec:comp} we discuss our model fits to the
data, the derivation of various physical quantities associates with the
X-ray emitting plasma, and how the SNR compares to other O-rich SNR.
Our results and conclusions are summarized in
\S \ref{sec:conc}.

\section{Observations and Data Reduction}
\label{sec:obs}

We have analysed an archival 29 ksec exposure of NGC 4449  
obtained on 4 Feb 2001 (ObsID No.2031; PI: T. Heckman) 
with the S3 chip of the Advanced CCD Imaging Spectrometer
(ACIS; \citealt{garmire92,bautz98}). ACIS-S3 is a back-side-illuminated CCD 
with low to moderate spectral resolution ($E/\Delta E \sim$ 4.3 at 0.5 keV;
$E/\Delta E$ $\sim$ 30 at 5.9 keV).
We used {\it Chandra X-Ray Observatory} center level 1 and
level 2 event lists, and the {\it ASCA}-like [0, 2, 3, 4, 6] grade
set.

\begin{figure}
\plotone{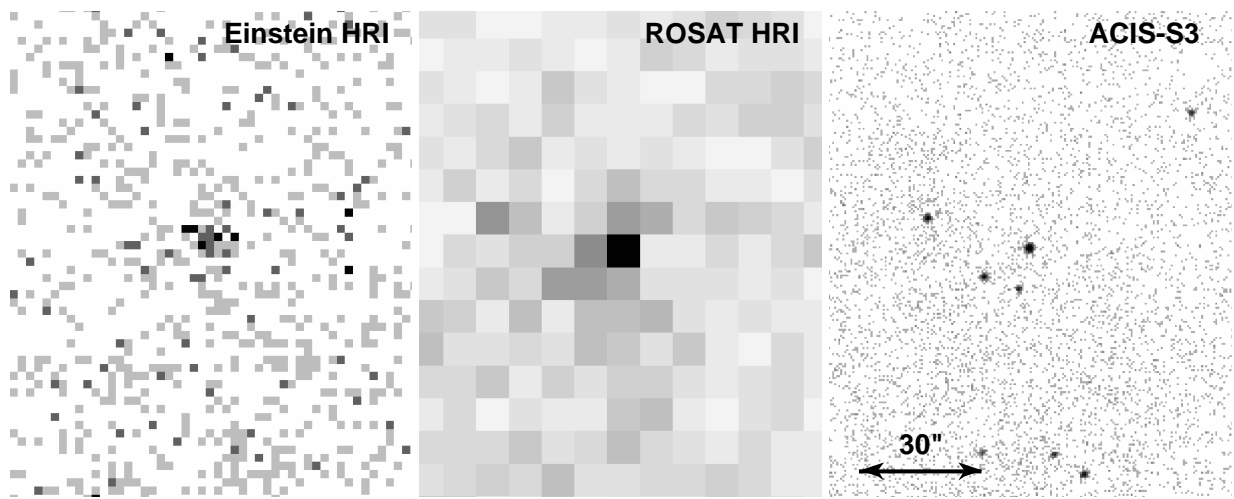}
\caption{Einstein HRI, ROSAT HRI, and Chandra ACIS-S images of the SNR in
NGC 4449.}
\label{fig:images}
\end{figure}

The {\it Chandra} ACIS image of the SNR is shown in Figure~\ref{fig:images}. 
For comparison, we also show
the Einstein HRI and ROSAT HRI images. In previous observations, it is 
clear that there is source confusion with two X-ray sources located 
$\approx$ 10\arcsec ~ to the southwest. However, the Chandra observation 
does not suffer from this source confusion, and we can separate the 
SNR from surrounding sources. We note that the position of the SNR is coincident
with the radio source as identified by \citet{seaquist78}.

The data were calibrated and background subtracted using the 
{\it Chandra Interactive Analysis of Observations} (CIAO) 
software (Version 2.2) along with the Version 2.15 of the calibration
database. 
A spectrum was extracted
from a circle of 10 pixel radius ($\approx$ 5\arcsec) centered on the
source.  While \citet{blair98} quote an upper limit of 0.6 pc ($0 \farcs 028$)
on the size of the remnant, the point spread function of the telescope
produces only 76\% of the encircled energy within a 1\arcsec~ diameter
aperture \citep{jerius00}. Given that the SNR is isolated with respect to 
other sources in NGC 4449, we chose an 
aperture of $\approx$ 10\arcsec, where 99\% of the source emission is
detected. A background spectrum located $\approx$ 15\arcsec ~ to the northeast
of the SNR was also extracted.

To help identify potential line features in the low S/N spectrum, 
we compared the
NGC 4449 SNR  spectrum to those of other O-rich SNRs.
Using the archival {\it Chandra} observation (ObsID No. 126) of G292.0+1.8, we
extracted the spectrum of the entire remnant. Because of variations in the
ACIS-S quantum efficiency, the $12' \times 8'$
remnant was binned into $0 \farcm 74 $ $\times$ $0 \farcm 74$ 
(90 $\times$ 90 pixels) bins.
The spectrum from each bin was extracted, and we summed all the
individual spectra. The summed spectrum was then convolved with an energy
and position weighted response matrix, as calculated by the CIAO task 
``{\it mkwarf~}''. A similar procedure was followed
to extract summed spectra for {\it Chandra} observations of 
Cas A (ObsID No. 114) and 1E 0102-7219 (ObsID No. 1231).

\section{Results and Discussion}
\label{sec:comp}

\begin{figure}
\plotone{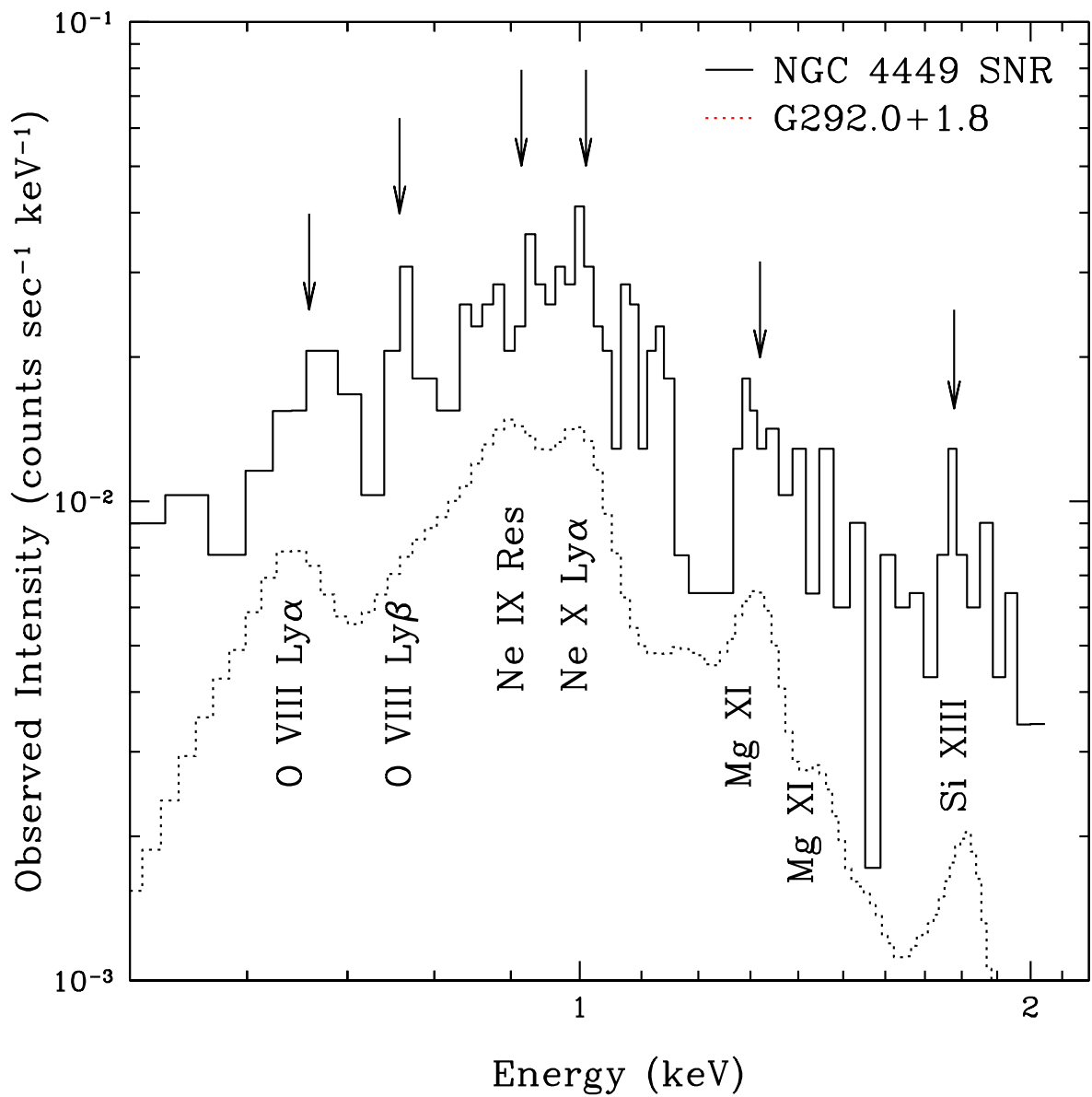}
\caption{Comparison between the SNR in NGC 4449 and SNR G292.0+1.8. The
observed intensity of G292.0+1.8 has been arbitrarily scaled down by a factor
of 8000 so that a comparison of detected lines between the two 
SNRs may be made. No correction for Galactic extinction has been included.}
\label{fig:4449}
\end{figure}

The {\it Chandra} ACIS spectrum of the NGC~4449 SNR for 0.5 to 2.1 keV is 
shown in Figure~\ref{fig:4449}. 
Using PI binning, the data were binned by a factor of four,
reducing the total number of pulse height channels from 4096 to 1024.
The observed background subtracted count rate was (3.8 $\pm$ 0.2) 
$\times$ 10$^{-2}$ counts s$^{-1}$. 
Although the full spectral range of ACIS is 0.1 to 9.0 keV, we have limited 
our analysis of the NGC~4449 SNR X-ray spectrum to between 0.5 to 2.1 keV.
Inspection of the data showed few detected events above 2.1 keV and below
500 eV.

Despite the low counts per channel ($<$ 25), 
five likely emission features can be seen 
in the extracted spectrum (see Fig.~\ref{fig:4449}). 
The clearest ones are what appear to be \ion{Ne}{10} at 1.05 keV and 
emission around the Si-K complex of lines near 1.85 keV. We also note the
possible detection of 
\ion{O}{8} at 654 and 774 eV, and   
\ion{Mg}{11} at 1.5 keV. 

For comparison, we also show in Figure~\ref{fig:4449} 
our {\it Chandra} ACIS summed spectrum of the Galactic O-rich SNR G292.0+1.8.  
One can see that there is a general resemblance of the NGC 4449 SNR spectrum to that
of G292.0+1.8 in that several 
emission features seem to be present in both and with similar strength (e.g., 
for example, \ion{O}{8}, \ion{Mg}{11}, and \ion{Si}{13} emission features.) 
Because of this resemblance,
we have used the {\it Chandra} G292.0+1.8 spectrum as a 
template to identify possible weak features in NGC 4449 SNR.
Specifically, G292.0+1.8 shows the presence of 
very weak \ion{Ne}{9} and \ion{Mg}{12}
emissions.  In spite of the relatively low quality of the NGC 4449 SNR spectrum,  
there are some indications for the presence 
of both lines, but with different relative strengths compared to SNR G292.0+1.8.
That is, the NGC 4449 SNR spectrum hints at weaker \ion{Ne}{9} emission 
relative to \ion{Ne}{10} than that seen in G292.0+1.8. 

\subsection{Spectral Fitting}

Because of the spectral similarity of the  NGC 4449 SNR with G292.0+1.8, 
we have employed the results of an {\it Einstein} analysis 
of G292.0+1.8 \citep{hughes94}
as a starting point for estimating the X-ray emission temperature and 
elemental abundances for the NGC 4449 SNR. 

\begin{figure}
\epsscale{0.8}
\plotone{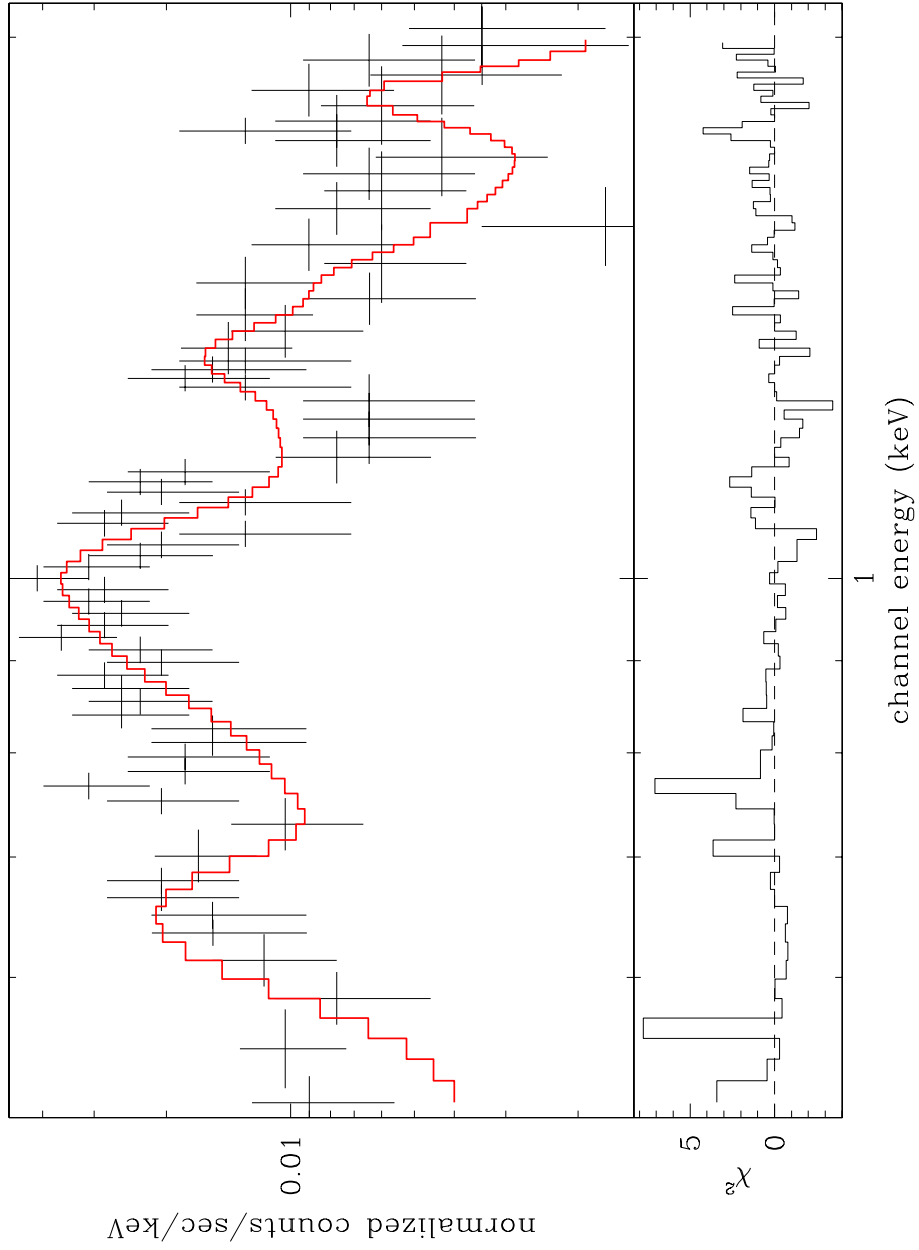}
\caption{Model fit to the spectrum extracted from the NGC 4449 SNR.}
\label{fig:model}
\end{figure}

The ACIS spectrum of the NGC 4449 SNR was fit to a non-equilibrium 
ionization (NEI) model with variable 
abundances, corrected for Galactic absorption. Due to the low quality of 
the data,
the spectral fits were overparametrized ($\chi^{2}_{red} \sim$ 0.6). 
However, a weighting scheme developed by \citet{churazov96}, whereby the
weight for a given data channel is estimated by averaging the counts in
surrounding channels, significantly improved the quality of the model fits
($\chi^{2}_{red} \sim$ 1.1). The resulting best fit is shown in 
Figure~\ref{fig:model}. 
Table~\ref{tab:param} lists our best fit values for the plasma 
temperature (kT), absorbing column density (N$_{\rm H}$), and 
ionization timescale ($n_et$). The quoted errors are at 90\% 
confidence for the 10 free parameters. For comparison, 
we also list the results of \citet{hughes94} for G292.0+1.8.
Our best fit values of
of N$_{\rm H}$ = 1.7 $\times$ 10$^{21}$ cm$^{-2}$ and
T $\approx$ 9.2 $\times$ 10$^{6}$
K, corresponding to a shock velocity of 830 km s$^{-1}$.
These compare favorably with the \citet{vogler97}  
values of N$_{\rm H} \geq $  0.7 $\times$ 10$^{21}$
and T $\geq$ 7 $\times$ 10$^{6}$ K. 
Our value for N$_{\rm H}$ also agrees well with 
the value derived by \citet{blair83}, 
N$_{\rm H}$  = 1.5 $\times$ 10$^{21}$ cm$^{-2}$
which was based on optical measurements.

Our resulting modeled $0.5 - 2.1$ keV X-ray flux for the NGC 4449 remnant 
is (8.6 $\pm$ 1.3) $\times$ 10$^{-14}$ erg cm$^{-2}$ s$^{-1}$, which 
at a distance of 3.9 Mpc, corresponds to 
L$_{x}$ = 2.4 $\times$ 10$^{38}$ erg s$^{-1}$. 
This is in reasonably good agreement with the value quoted by \citet{vogler97} of
L$_{x}$ $\approx$ 1.8 $\times$ 10$^{38}$ erg s$^{-1}$.
However, it is about half the  5 $\times$ 10$^{38}$
erg s$^{-1}$ value (at a revised distance of 3.9 Mpc) determined from
{\it Einstein} HRI measurements by \citet{blair83}.
While the cause for this discrepancy may be due to a real
decrease in X-ray flux (see Section 3.3 below), the {\it Einstein} HRI measurement was based
on data covering a wider energy band ($0.2 - 4$ keV) and 
was modeled with a Raymond-Smith plasma assuming solar abundances.

Although the ACIS spectrum of the NGC 4449 SNR is of low S/N, 
it is still possible to estimate some physical parameters of 
the remnant's X-ray emitting plasma. We first estimated  
the mass of ejecta which is currently emitting in 
X-rays.  We began by assuming a simple geometry
for the remnant's X-ray emission structure and 
then use a spectral model, such as our NEI model, 
to relate the observed X-ray emissivity to the volume 
emission measure.  Given that the source is unresolved, 
we assumed a spherical geometry. 
Then, adopting a remnant radius of 0.3 pc \citep{blair98}, 
the total volume of the emitting
region is  V = 3.1 $\times$ 10$^{54}$ {\it f} D$^{3}_{3.9 Mpc}$ cm$^{3}$, 
where D$_{3.9 Mpc}$ is the distance to the remnant, 
normalized to 3.9 Mpc and
{\it f} is
the volume filling factor. The root mean square hydrogen number density is
determined from the model normalization, which is just the volume emission
measure, $n_{e}$$n_{\rm H}$V, for a known distance and geometry. 

From Table~\ref{tab:param},
$n_{e}$$n_{\rm H}$V = (5.5 $\pm$ 0.05) $\times$ 10$^{60}$ cm$^{-3}$. 
Using 
n$_e$/n$_{\rm H}$ = 1.47, consistent with the ratio used by
\citet{hughes94}, and the expression for the emitting volume above,
the root mean square hydrogen number density is n$_{\rm H}$ =
1.1 $\times$ 10$^3$ {\it f}$^{-0.5}$ D$^{-0.5}_{\rm 3.9 Mpc}$
cm$^{-3}$.  Using this expression, we estimate the mass of the X-ray 
emitting gas, including the mass of metals as well as He and H, to be  
M = 5.7 {\it f} $^{0.5}$ D$^{2.5}_{\rm 3.9 Mpc}$ M$_{\sun}$.

We have also used the {\it Chandra} spectrum to estimate elemental
abundances within the remnant ejecta. In 
Table~\ref{tab:param}, we list the volumetric emission measure for each
element we fit, scaled to solar values. By again assuming an 
$n_e$/$n_{\rm H}$ of
1.47, we determined the relative abundances for each
element listed. For a particular element, the abundance relative 
to solar is given by:

\begin{displaymath}
\frac{{\rm (X/H)}}{\;\;{\rm (X/H)}_{\sun}}  = 
n_en_{\rm H}{\rm V} \frac{n_{\rm X}/n_{\rm H}}{{\rm (X/H)}_{\sun}} \; \, .
\end{displaymath}

\noindent
Consequently, our estimated abundances for 
O, Ne, Mg, Si, and Fe are 18, 15, 9.0, 9.0, and 2.2 by 
number, respectively.
The high overabundance of oxygen is consistent with the 
remnant's UV and optical spectrum 
which is dominated by strong oxygen line emissions \citep{kirshner80,blair84}.

\subsection{Comparisons to Other Oxygen-Rich SNRs}

X-ray and optical properties for
five of the seven currently known O-rich SNRs are listed in 
Table~\ref{tab:orich}. 
For each remnant other than the one in NGC 4449, unabsorbed fluxes and
luminosities are listed for the energy range 0.3 -- 2.1 keV. These values 
were obtained by fitting a two component thermal plasma with variable 
abundances to the total SNR spectra 
\citep{seward02}. 

With an X-ray luminosity of 2.4 $\times$ 10$^{38}$ erg s$^{-1}$,
the NGC 4449 SNR
is clearly the most luminous O-rich SNR
relative to the the four other O-rich SNRs
listed in Table~\ref{tab:orich}, which includes Cas A.
The origin of this high luminosity is most likely a strong shock 
interaction with dense local interstellar and/or circumstellar material (CSM).
However, the young age of the NGC 4449 SNR may also be an important factor.
Similar and even much higher X-ray luminosities have been seen
in some historic extragalactic SNe having strong CSM interactions some 10 to 30 years after outburst 
(e.g., SN 1979C:
14 $\times$ 10$^{38}$ erg s$^{-1}$; SN 1995N:
175 $\times$ 10$^{38}$ erg s$^{-1}$; for 0.1-2.4 keV).

Recent optical spectral line data suggest expansion
velocities in the remnant as high as 6000 km s$^{-1}$ \citep{blair98}.
Such velocities, when combined with an upper limit for the angular diameter of 0.6 pc
from HST observations, imply an age of the SNR is $\approx$
100 yr. This makes the NGC 4449 SNR only about one tenth
as old as G292.0+1.8, N132D, or 1E 0102-7219, and about a third as 
old as Cas A, the youngest currently known Galactic SNR.  

\begin{figure}
\plotone{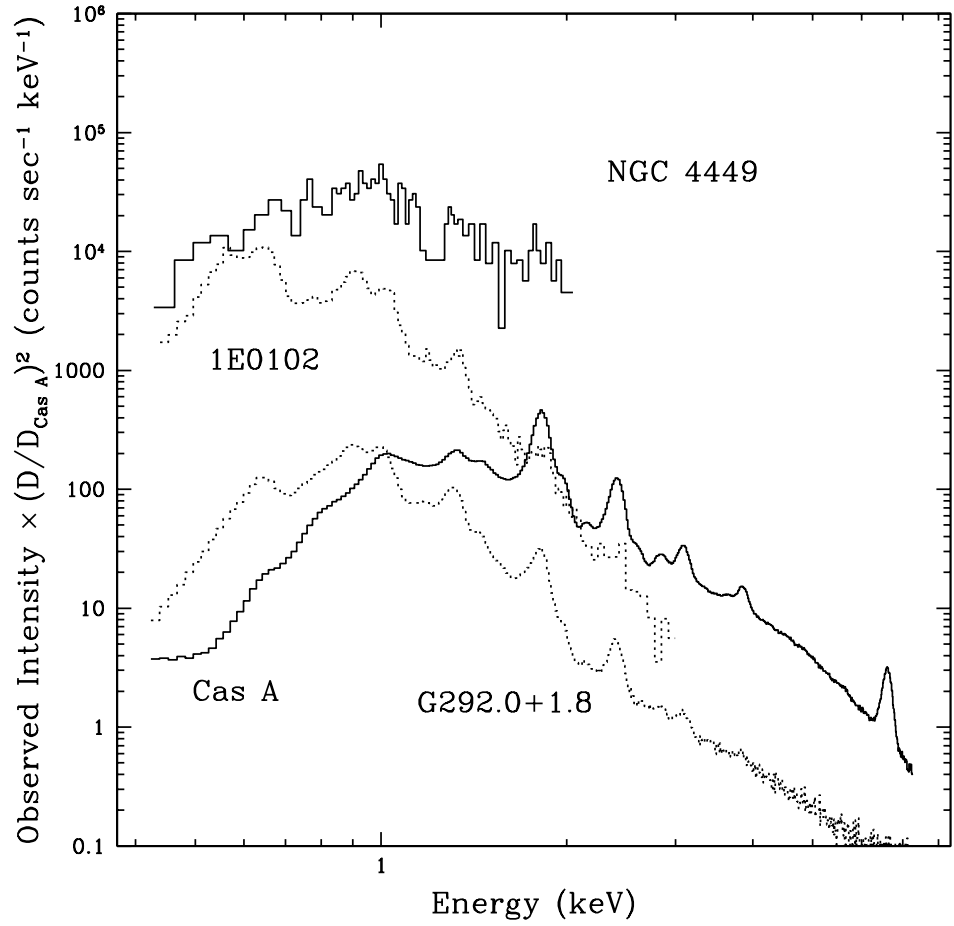}
\caption{Spectra for four oxygen-rich SNR. Each spectrum is for the
whole remnant, scaled to a distance of 3.4 kpc. No corrections for 
Galactic extinction have been made.}
\label{fig:orich}
\end{figure}

We show in Figure~\ref{fig:orich} the extracted NGC 4449 SNR spectra 
compared to the O-rich SNRs Cas A, 1E~0102-7219, and G292.0+1.8. 
Significant differences can be seen among these four SNR spectra. 
For example, while G292.0+1.8, 1E 0102-7219, and the NGC 4449 SNR all 
show emission around the 900 eV \ion{Ne}{9} resonance forest 
(see Fig.~\ref{fig:4449}), this emission
is nearly absent in Cas A. 
Also, 1E 0102-7219 is the only SNR
shown which exhibits obvious emission at the 574 eV \ion{O}{7} line.
Therefore, while there is clearly a range of X-ray spectral properties
within the O-rich subclass of SNRs, the X-ray spectrum of the
NGC 4449 SNR fits within this range.

Except for G292.0+1.8, there are large differences in the expansion 
velocity derived from X-rays and that derived
from optical emission lines, as shown in Table~\ref{tab:orich}.
The shock velocity estimate for the X-ray emitting gas in
NGC 4449 SNR is 830 km s$^{-1}$. This is $\approx$ 15\% higher than that
estimated from the spectral fit of \citet{vogler97} 
($\approx$ 700 km s$^{-1}$), but
still many times lower than
the 3500 -- 6000 km s$^{-1}$ velocity measured from optical line emission
\citep{blair83,blair98}.
Similar differences between X-ray and optical 
expansion velocities have been seen in other O-rich
SNR. Optical studies of Cas A 
indicate that the ejecta is moving at 4000 -- 6000 km s$^{-1}$. 
However, X-ray analyses suggest that the X-ray emitting gas has a 
temperature of $\sim$ 3.0 $\times$ 10$^7$ K, corresponding to a 
velocity of $\approx$ 1900 km s$^{-1}$. Similarly, in N132D and 
1E 0102-7219 the X-ray shock velocity is only about 20\% that of
the velocity seen from optical emission studies.

The fact that the X-ray emitting gas exhibits a lower shock velocity
than the optically emitting gas is not unexpected. Both optical and  
X-ray emissions largely arise from SN ejecta interacting with the remnant's reverse shock 
that is generated as the
SNR forward shock moves through local ISM/CSM material. The reverse shock,
traveling back toward the SNR center, will naturally drive a lower velocity shock
into the denser optically bright ejecta compared to the lower density, X-ray gas.
The higher reverse shock velocity in the X-ray bright gas will lead to greater
deceleration of the outward expanding ejecta.
Thus, the remnant's X-ray derived expansion will be corresponding less 
than that seen in the optical.

An additional factor for explaining the difference between the optically 
derived expansion velocity and the X-ray gas velocity is that in all cases, 
equipartition between ion and electron temperatures has been assumed in the 
X-ray spectral model fits. In general the ions should have a higher 
temperature, since they carry the bulk of the energy in the shock. If the 
electrons and ions are not equilibrated, which is possible in collisionless 
shocks like those found in SNR where equilibrium is reached via Coulomb 
collisions, then the mean gas temperature should be more heavily weighted 
toward the ion temperature, implying a X-ray shock velocity closer to the 
shock velocity seen in optical emissions. Recently, \citet{ghavamian02} have 
found a very low degree of electron-ion equilibration ($T_e/T_p$ $\leq$ 0.07) 
in the SN 1006 remnant. If this sort of non-equilibrium state
existed in the NGC 
4449 SNR, as well as many of the the other O-rich SNR listed, then that might 
help to explain the relatively low derived  X-ray shock velocities. 

\subsection{SNR Evolution}

\citet{kirshner80} suggested that because of its young age  
and prodigious luminosity,
significant changes in the X-ray and optical output of the NGC 4449 SN might 
be observed on the timescale of just a few years.   
The 2001 {\it Chandra} flux reported here may
lend support to growing evidence for relatively rapid changes
in the remnant's properties and luminosities.

For example, \citet{debruyn81} and \citet{debruyn83} observed a nearly 50\%
drop in the 6 cm radio flux density between 1973 and 1982.
In the optical, whereas 1978 -- 1980 spectra taken by 
\citet{kirshner80} and \citet{blair83} found only narrow [\ion{S}{2}]
$\lambda \lambda$6716,6731 associated with
emission from the surrounding \ion{H}{2} region, recent spectra
obtained in 1995 and 2002 indicate the 
emergence of broad [\ion{S}{2}] $\lambda \lambda$6716,6731
from the SN ejecta \citep{fesen03}.

In light of these results, our X-ray flux measurement 
may likewise indicate 
changes in the remnant's X-ray flux over the last 20 years.
Both our 2001 {\it Chandra} and the 1994
{\it ROSAT} estimated $0.5 - 2.1$ keV  X-ray luminosity values are  only about
half of the 1979 {\it Einstein} $0.2 - 4.0$ keV measurements reported  
by \citet{blair83}. Although the energy ranges for these flux measurements
are different, the observed {\it Chandra} spectrum does not suggest a significant amount of
flux below 0.5 keV or above 2.1 keV. Thus, while there is some uncertainty 
in the accuracy of comparing {\it Einstein} and {\it Chandra} derived 
fluxes for such a metal-rich SNR, the overall percentage decrease observed 
is not out of line with reported changes seen in other wavelength regions. 

In order to investigate further the reality of these flux changes, 
we have convolved our NGC 4449 SNR X-ray spectrum 
model with the telescope response for both the {\it Einstein} HRI and {\it ROSAT} HRI.
This resulted in a predicted count rate which then could be directly compared to the
observed {\it Einstein} and {\it ROSAT} count rates for the SNR
reported in the literature. The reported {\it Einstein} and {\it ROSAT} 
count rates are most likely overestimated
due to source confusion from two nearby X-ray sources (see Fig.~\ref{fig:images}). 
To account for this, we
assumed that the two nearby sources were stable over the time period 1979 -- 2001. 
We calculated the ratio of the  
{\it Chandra} count rates for the two sources to the {\it Chandra} count rate for the SNR plus the contaminating
sources. This percentage was then subtracted from the observed {\it Einstein} and {\it ROSAT} count rates to
get an adjusted rate. This method assumes that the SNR count rate was also stable between
the different observations. Finally, in order to remove telescope dependent effects, we
adopted the ratio of the source-adjusted observed rate to the predicted rate.

\begin{figure}
\plotone{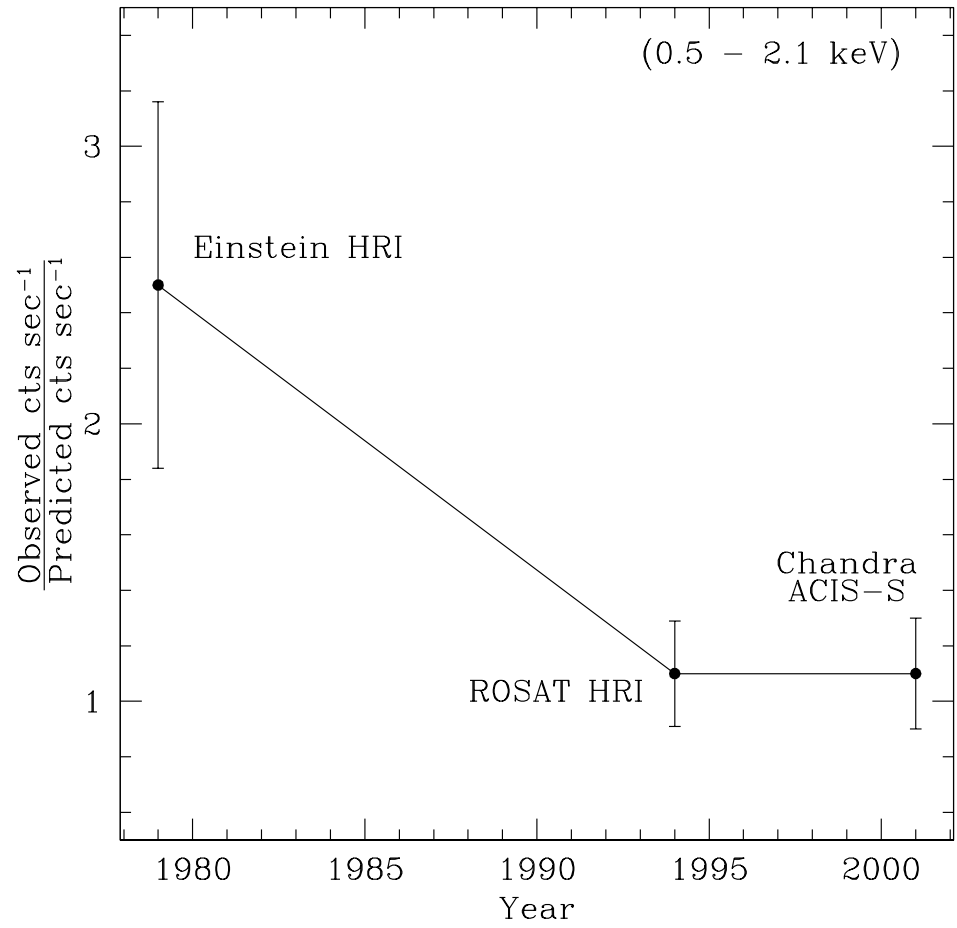}
\caption{Ratio of the observed count rates over three observations versus 
count rates predicted by convolving our NEI model with the telescope 
responses.}
\label{fig:ratio}
\end{figure}

The ratio of source-adjusted observed to predicted
count rate is shown in Figure~\ref{fig:ratio}. 
In essence, this plot shows the ratio of the observed count rate versus what would be observed
by the same telescope in 2001. This figure suggests that the count rate, and
thus the remnant's X-ray luminosity, has dropped nearly 50\% over the past 
15--20 years. It will be interesting to follow the remnant's X-ray
evolution over the next few decades to see if this trend continues.

\section{Conclusions}
\label{sec:conc}

We have analyzed a 29 ksec {\it Chandra} ACIS-S
observation of the  $\approx$ 100 yr old O-rich supernova remnant in the
irregular galaxy NGC 4449.
We found the following:

1) The spectrum shows the
likely presence of several emission lines including \ion{O}{8}, \ion{Ne}{10},
\ion{Mg}{11}, and \ion{Si}{13} with an overall spectrum
similar in appearance to the oxygen-rich Galactic SNR G292.0+1.8.

2) We find an X-ray luminosity of 2.4 $\times$ 10$^{38}$ erg s$^{-1}$,
for the energy range 0.5 -- 2.1 keV and a distance of 3.9 Mpc. 
Assuming a spherical geometry for the remnant
we estimate the mass of the X-ray
emitting gas, including the mass of metals as well as He and H, to be
M$_{\rm Xray}$ $\approx$ 5.7 {\it f} D$^{2.5}_{\rm 3.9 Mpc}$ M$_{\sun}$.

3) A non-equilibrium ionization
fit to the spectrum suggests a large overabundance of oxygen of $\approx$ 18 times
solar, consistent with the remnant's UV and optical emission-line properties.
 
4) Based on electron-ion temperature equilibration, we derive an 
X-ray shock velocity of $\approx$ 830 km s$^{-1}$. 
This is significantly lower than the 6000 km s$^{-1}$ 
optical expansion velocity determined from optical measurements but 
consistent with differences between the optical and X-ray velocities seen
in other O-rich remnants.

5) While our estimated  X-ray flux is in rough agreement with that
measured by {\it ROSAT} in 1994, it is only about half of that
of an 1978 {\it Einstein} observation. Despite uncertainties in the
accuracy of such flux measurement comparisons, such a decrease 
would not be unexpected given recent measurements
made in the radio and optical suggesting that the remnant continues 
to undergo significant changes since it discovery in 1978.

The NGC 4449 SNR's current standing as the youngest and most luminous O-rich remnant
known certainly make it an object worthy of continued study. 
The remnant's estimated angular dimensions of $\approx$ $0 \farcs 028$ make it  
an ideal candidate for VLBI measurements of its true size and emission sub-structure,
like that accomplished for SN 1993J \citep{bietenholz01}.
Such measurements could help clarify its general structure and,  
when combined with optical emission line width data, 
set strict limits on its expansion age.   

\acknowledgements

The authors wish to thank Robert Petre and Randall Smith for useful 
comments during the preparation of this manuscript,
Dick Edgar for help with fitting the data, and
John Raymond for helpful discussions regarding the effects of 
non-equipartition in the plasma temperature. We also thank the anonymous
referee for useful suggestions and a careful reading of this manuscript.

\begin{deluxetable}{lrr}
\tablecolumns{3}
\tablecaption{Best-Fit Parameters for NEI Model}
\tablewidth{0pc}
\tablehead{
\colhead{Parameter} & \colhead{NGC 4449\tablenotemark{a}} & \colhead{G292.0+1.8\tablenotemark{a,b}}}
\startdata
N$_H$ (atoms cm$^{-2}$) \dotfill & (1.7 $\pm$ 0.06) $\times$ 10$^{21}$ & (8.0$^{+0.68}_{-1.98}$) $\times$ 10$^{21}$ \\
kT (keV) \dotfill & 0.80$^{+0.28}_{-0.20}$ &  1.64$^{+0.29}_{-0.19}$ \\
n$_e${\it t} (s cm$^{-3}$) \dotfill & (1.18 $\pm$ 0.04) $\times$ 10$^{12}$ & (5.55$^{+1.2}_{-1.1}$) $\times$ 10$^{10}$ \\
$<$kT$>$ (keV) \dotfill & 1.19$^{+0.09}_{-0.07}$  & \nodata \\
n$_{\rm H}$n$_e$V (cm$^{-3}$)\tablenotemark{c} \dotfill & (0.55 $\pm$ 0.05) $\times$ 10$^{61}$ & (0.43 $\pm$ 0.37) $\times$ 10$^{58}$ \\
n$_{\rm O}$n$_e$V/[O/H]$_{\odot}$ (cm$^{-3}$) \dotfill & (10.0$^{+6.7}_{-3.8}$) $\times$ 10$^{61}$ & (22.6$^{+2.7}_{-3.2}$) $\times$ 10$^{58}$ \\
n$_{\rm Ne}$n$_e$V/[Ne/H]$_{\odot}$ (cm$^{-3}$) \dotfill & (2.0$^{+1.5}_{-1.4}$) $\times$ 10$^{61}$ & (6.4$^{+2.7}_{-3.2}$) $\times$ 10$^{58}$\\
n$_{\rm Mg}$n$_e$V/[Mg/H]$_{\odot}$ (cm$^{-3}$) \dotfill & (4.8$^{+2.8}_{-2.7}$) $\times$ 10$^{61}$ & (7.88 $\pm$ 0.40) $\times$ 10$^{58}$ \\
n$_{\rm Si}$n$_e$V/[Si/H]$_{\odot}$ (cm$^{-3}$) \dotfill & (5.1$^{+1.7}_{-1.6}$) $\times$ 10$^{61}$ & (2.14 $\pm$ 0.14) $\times$ 10$^{58}$ \\
n$_{\rm Fe}$n$_e$V/[Fe/H]$_{\odot}$ (cm$^{-3}$) \dotfill & (1.2$^{+0.3}_{-0.3}$) $\times$ 10$^{61}$ & (3.0$^{+1.1}_{-1.0}$) $\times$ 10$^{58}$ \\

[O/H]/[O/H]$_{\odot}$ \dotfill & 18.0$^{+12.0}_{-6.5}$ & \nodata \\

[Ne/H]/[Ne/H]$_{\odot}$ \dotfill & 15.0$^{+12.0}_{-10.0}$ & \nodata \\

[Mg/H]/[Mg/H]$_{\odot}$ \dotfill & 9.0$^{+5.5}_{-5.0}$ & \nodata \\

[Si/H]/[Si/H]$_{\odot}$ \dotfill & 9.0$^{+3.1}_{-3.0}$ & \nodata \\

[Fe/H]/[Fe/H]$_{\odot}$ \dotfill & 2.2$^{+0.6}_{-0.5}$ & \nodata \\
$\chi^2$ (d.o.f) \dotfill & 292 (266) & 154 (130) \\
\enddata
\tablenotetext{a}{Statistical errors at 90\% confidence.}
\tablenotetext{b}{Values quoted from \citealt{hughes94}.}
\tablenotetext{c}{Emission measure scaled by solar abundance relative to H. 
Assumes a distance of 3.9 Mpc to NGC 4449, and SNR radius of 0.3 pc. 
For the best fit values of the parameters in the table, 
$n_e/n_{\rm H}$ = 1.47.}
\label{tab:param}
\end{deluxetable}

\begin{deluxetable}{lrccrccc}
\tablecolumns{10}
\tabletypesize{\scriptsize}
\tablecaption{Comparison of O-rich SNR}
\tablewidth{0pc}
\tablehead{
\colhead{Object} & \colhead{Distance} & \colhead{F$_{\rm X} $\tablenotemark{a}}
& \colhead{L$_{\rm X}$} & \colhead{Age} & \colhead{V$_{\rm opt}$} 
& \colhead{V$_{\rm X-ray}$} & \colhead{} \\
\colhead{} & \colhead{(kpc)} & 
\colhead{(10$^{-14}$ erg cm$^{-2}$ s$^{-1}$)} & 
\colhead{(10$^{38}$ erg s$^{-1}$)} & 
\colhead{(yr)} & \colhead{(km s$^{-1}$)} & \colhead{(km s$^{-1}$)}
& \colhead{References}}
\startdata
NGC 4449 SNR\tablenotemark{b} & 3900 & 8.6 & 2.4 & $\sim$ 100 & 6000 & 830 & 1,2,3,4\\
G292.0+1.8\tablenotemark{c} & 4.8 & 2.0 $\times$ 10$^{5}$ & 0.06 & $<$ 1600 & 1100 & 1200 & 5,6,7 \\
Cas A\tablenotemark{c} & 3.4 & 1.9 $\times$ 10$^{6}$ & .26 & 320 & 6000 & 1900 & 6,7 \\
N132D\tablenotemark{c} & 49 & 3.3 $\times$ 10$^{4}$ & 0.9 & $\sim$ 1350 & 3400 & 790 & 6,7 \\
1E 0102.1-7219\tablenotemark{c} & 72 & 5.8 $\times$ 10$^{3}$ & 3.4 $\times$ 10$^{-3}$ & $\sim$ 1000 & 4000 & 610 & 6,7 \\
\enddata
\tablenotetext{a}{Unabsorbed flux determined from fitting a two component
thermal plasma with variable abundances.}
\tablenotetext{b}{For the energy range 0.5 -- 2.1 keV.}
\tablenotetext{c}{For the energy range 0.3 -- 2.0 keV.}
\tablerefs{(1) \citealt{balick78}; (2) \citealt{seaquist78}; 
 (3) \citealt{blair83}; (4) \citealt{blair98}; 
(5) \citealt{hughes94}; (6) \citealt{seward02}; (7) \citealt{vandenbergh88}}
\label{tab:orich}
\end{deluxetable}

\end{document}